\documentclass{article}

\usepackage{amsmath, amssymb, amsfonts, fullpage}
\usepackage[english]{babel}
\usepackage{biblatex}
\usepackage[titletoc,title]{appendix}

\newtheorem{assumption}{Assumption}
\usepackage{xcolor}
\newcommand{\E}{\mathbb{E}}
\newcommand{\I}{\text{I}}

\newcommand\indep{\protect\mathpalette{\protect\independenT}{\perp}}
\def\independenT#1#2{\mathrel{\rlap{$#1#2$}\mkern2mu{#1#2}}}

\providecommand{\keywords}[1]
{
  \small	
  \textbf{\textit{Keywords: }} #1
}

\providecommand{\coi}[1]
{
  \small	
  \textbf{\textit{Conflicts of interest: }} #1
}
\providecommand{\funding}[1]{
  \small	
  \textbf{\textit{Sources of funding: }} #1
}
\providecommand{\publicationhistory}[1]{
  \small	
  \textbf{\textit{Publication history: }} #1
}

\title{Pulling back the curtain: the road from statistical estimand to machine-learning based estimator for epidemiologists (no wizard required)}
\author{Audrey Renson$^1$, Lina Montoya$^{2,3}$, Dana E. Goin$^4$, Iv\'an D\'iaz$^1$, Rachael K. Ross$^4$\\
\\
\small $^1$Department of Population Health, New York University Grossman School of Medicine, New York, United States\\%
\small $^2$School of Data Science and Society, University of North Carolina at Chapel Hill\\%
\small $^3$Department of Biostatistics, Gillings School of Global Public Health, University of North Carolina at Chapel Hill\\%
\small $^4$Department of Epidemiology, Mailman School of Public Health, Columbia University,
New York, United States}
\date{
    \today
}

\begin{document}

\maketitle

\begin{abstract}
Epidemiologists increasingly use causal inference methods that rely on machine learning, as these approaches can relax unnecessary model specification assumptions. While deriving and studying asymptotic properties of such estimators is a task usually associated with statisticians, it is useful for epidemiologists to understand the steps involved, as epidemiologists are often at the forefront of defining important new research questions and translating them into new parameters to be estimated. In this paper, our goal was to provide a relatively accessible guide through the process of (i) deriving an estimator based on the so-called efficient influence function (which we define and explain), and (ii) showing such an estimator's ability to validly incorporate machine learning, by demonstrating the so-called rate double robustness property. The derivations in this paper rely mainly on algebra and some foundational results from statistical inference, which are explained. 
\end{abstract}
\\
\\
\small \textbf{\textit{Word count: }} $\approx$4,086 / 4,000\\
\\
\vspace{1cm}
\\
\coi{None to disclose.}\\
\\
\funding{AR is supported by a gift from the Bezos foundation. DEG is supported by award number R00ES033274 from the National Institute of Environmental Health Sciences. LMM is supported by the National Institutes of Health under award number R00MH133985.}\\
\\
\publicationhistory{Publication history: This manuscript was previously published in arXiv: doi:}
\newpage

\section{Introduction}
Methods in causal inference that rely on machine learning to estimate nuisance functions (e.g., outcome regressions, propensity scores) are increasingly popular because they can avoid modeling assumptions not supported by scientific knowledge. Such methods are usually well-described (and implemented in software) when the causal parameter and set of assumptions used to identify it are of a familiar variety, such as the average treatment effect (ATE) under conditional exchangeability, positivity, and consistency \cite{van2011targeted}. However, if we want to estimate an uncommon causal parameter in order to answer our research question, or if causal identification is based on a less common set of assumptions, estimators may not yet be available. Examples of such uncommon parameters from our own research include the time-varying g-formula under parallel trends \cite{illenberger2024efficient}, transported effects from difference-in-differences \cite{, renson2023transporting}, and transported measurement error corrections \cite{ross2024leveraging}.

The key challenge in deriving estimators that allow machine learning is in obtaining valid inference (e.g.,confidence intervals). For example, it is understood that doubly robust estimators of the ATE can allow calculation of confidence intervals with nominal coverage (in large samples) even when machine learning is used, while the g-computation estimator cannot outside of special cases \cite{naimi2023challenges,balzer2023invited}. Estimators that allow valid inference under machine-learning-based estimation of nuisance functions are usually based on the efficient influence function (EIF), which we define and discuss below. 

The process for deriving EIF-based estimators and determining their properties has historically been limited to a specialized and highly technical branch of statistical literature. While deriving and studying such estimators in a mathematically rigorous way does require specialized training, in our experience substantial progress can be achieved with less mathematical skill than that required to read and understand most of the literature on this topic. Epidemiologists are often at the forefront of defining new research questions, translating them into causal parameters, and developing reasonable assumptions to identify these parameters; thus, epidemiologists can benefit from understanding the steps to derive machine-learning based estimators. While many reviews exist on the theory and application of this approach \cite{van2011targeted, hines2022demystifying, diaz2020machine, kennedy2022semiparametric}, there remains a gap in translation to a more applied audience. In this paper, we aim to provide a more accessible guide through the steps needed to propose an EIF-based estimator for a unique statistical estimand that corresponds to a relevant causal question, and verify that our resulting estimator can indeed accommodate machine learning.

\section{Step 1: Define the statistical estimand}\label{sec:estimand}
Throughout this paper we work with an example estimand motivated by the hypothetical research question, ``What would be the average outcome in our sample if everyone were untreated?'' 

We begin by translating the scientific question to a causal parameter. Let $Y$ denote the observed outcome, $A$ a binary treatment, and $W$ some set of observed covariates, so that the observed data can be denoted $O_i=(W_i,A_i,Y_i)$ for $i=1,...,n$, where $n$ denotes the sample size. We assume the data are independent and identically distributed (iid), though the approaches we present can be extended to non-iid data. We use uppercase to refer to observed values of random variables, and lower case for specific values. Let $Y^a$ denote the potential outcome when $A$ is set to $a$. Using this notation, we can translate the research question to a causal parameter $\E(Y^0)$, the expected outcome under no treatment. Note that this causal parameter is one piece of the the average treatment effect, $\E(Y^1) - \E(Y^0)$. 

Then, we use causal assumptions to equate the causal parameter to a statistical estimand based only on the observed data distribution (i.e., to achieve identification). For our example, the causal parameter is identified under three assumptions: 1) conditional exchangeability, $Y^0 \indep A|W$; 2) positivity, $f(w)>0 \implies \Pr(A=0|W=w)>0$ (where $f$ denotes a probability density or mass function); and 3) causal consistency, $A=a \implies Y=Y^a$. Conditional exchangeability can be assessed with the help of a causal diagram based on subject matter knowledge \cite{greenland1999causal, robins2001data}. Under these assumptions, it follows that $\E(Y^0)=\E(\E(Y|W,A=0))$; the latter is our statistical estimand which we will denote $\psi=\E(\E(Y|W,A=0))$. 

It is often useful to think of the statistical estimand as a \textit{mapping}; i.e., a function $\psi(P)$ whose input is a distribution $P$ and whose output is a scalar. That is, if the observed data $O$ is drawn from some distribution $P$, we can write $\psi(P)\equiv \psi \equiv \E_P(\E_P(Y|W,A=0))$, where the subscripts indicate that the expectation is taken with respect to $P$. 

As in much of the literature on machine-learning-based estimators in causal inference, we focus on the statistical estimand $\psi=\E(\E(Y|W,A=0))$ because it offers a simple introduction to the material presented and is commonly used in applied research. However, the need to derive estimators usually arises because our scientific question or identification result lead to a less-commonly considered estimand. To aid in adapting this approach to more complex problems, we illustrate in Appendix \ref{app:alt} the derivations for another causal parameter, the expected outcome under no treatment \textit{among the treated}, one piece of the average treatment effect in the treated.

\section{Step 2: Derive the EIF}
\subsection{What is the EIF and why do we care?}
We define the EIF formally in Appendix \ref{app:eif}; here, we give a heuristic description for intuition. The EIF is a type of derivative: if the statistical estimand is a function of a distribution, then the EIF is a derivative of that function with respect to the distribution, and thus captures how sensitive the estimand is to small changes in the data distribution. The EIF can also be understood as describing the \emph{influence} that removing unit $i$'s data would have on any efficient estimator \cite{hampel1974influence}. Critically for our purposes, the EIF provides a recipe for constructing machine-learning based estimators that allow valid inference. Specifically, the EIF describes the ``first order error'' of an estimator (defined below); this error is what generally invalidates inference when machine learning is used, and is thus essential to account for. 

For an estimand $\psi$, we denote its EIF as $\phi(O_i, P)$, emphasizing that it is a function of both the unit-level data and the data distribution. Importantly, EIFs are always constructed such that they have mean zero at the true distribution $P$. For example, the EIF for the mean of $Y$, $\E[Y]$, is  $\phi(O_i, P)=Y_i-\E[Y].$  As another useful example, the EIF of a conditional mean $\E[Y|X=x]$ for discrete $X$ is $\phi(O_i, P)=\frac{I(X_i=x)}{\Pr(X=x)}\{Y_i-\E[Y|X=x]\}$, where $I(\cdot)$ is the indicator function that returns 1 if its argument is true and 0 otherwise.

\subsection{Deriving the (conjectured) EIF}

We now illustrate a method to propose the EIF of an estimand. Deriving EIFs is a complex subject and many methods have been developed. Here, we present a relatively simple method that has previously been described \cite{van2011targeted,kennedy2022semiparametric}. Importantly, this method only conjectures (but does not prove) that a given function is the EIF (discussed  more below). Note also that throughout we assume data come from a nonparametric statistical model (i.e. no restrictions are placed on the observed data distribution), so that all EIFs presented are for nonparametric models (calculations of EIFs for semiparametric models are also possible, but outside our scope).  

The method uses three tools. First, we pretend that the data include only discrete variables, allowing us to express our estimand as a summation, $\psi=\E(\E(Y|W,A=0))=\sum_w \E(Y|W=w,A=0) \Pr(W=w).$ Second, we use the fact that the EIF is a derivative, and use derivative rules (e.g., sum and product rules) to convert the EIF of the estimand into an algebraic combination of EIFs of simpler parts (such as conditional expectations). We use the notation $EIF(\psi)$ to denote the operator that returns the EIF of a statistical estimand $\psi$. The sum rule states that the EIF of the sum of parameters is the sum of their respective EIFs, and the product rule states that for a parameter $\psi=ab$, its EIF is given by $EIF(ab)=EIF(a)b + EIF(b)a$. Third, these simpler parts usually have EIFs which have been derived and are well-known; we substitute these known EIFs into the expression and simplify. In particular, here (as is common) we use the EIFs for marginal and conditional means provided in the previous subsection.  

Beginning with the first step, we pretend $W$ is discrete and rewrite our estimand $\psi = \sum_w q(w) p(w)$ where $q(w)=\E(Y|W=w,A=0)$ and $p(w)=\Pr(W=w)$. Then applying the derivative rules, we have:
    \begin{align*}
        EIF(\psi)&=EIF\left\{ \sum_w q(w) p(w) \right\} \\
        &= \sum_w EIF\bigg\{q(w) p(w)\bigg\} \\
         &= \sum_w EIF\bigg\{q(w)\bigg\} p(w) + \sum_w q(w) EIF\bigg\{p(w)\bigg\}, 
    \end{align*}
where we use the sum and product rules in the second and third equality, respectively. Finally, we substitute known EIFs for sub-parts; specifically, $EIF\{q(w)\}=\frac{I(W=w,A=0)}{\Pr(W=w,A=0)}\left\{Y-q(w)\right\}$ and $EIF\{p(w)\}=\I(W=w)-p(w)$ (since $\Pr(W=w)$ can also be written as $\E[ I(W=w) ]$ and then we can substitute the EIF of a mean). We have:
 \begin{align*}
        EIF(\psi)&=\sum_w \frac{I(W=w,A=0)}{\Pr(W=w,A=0)}\bigg\{Y-q(w)\bigg\}p(w) + \sum_w q(w) \bigg\{I(W=w)-p(w)\bigg\} \\
         &=\sum_w \frac{I(W=w,A=0)}{\Pr(A=0|W=w)}\bigg\{Y-q(w)\bigg\} + \sum_w q(w) \bigg\{I(W=w)-p(w)\bigg\}, 
\end{align*}
where we apply Bayes' rule (i.e., $\frac{Pr(W=w)}{Pr(W=w,A=0)} = \frac{1}{Pr(A=0|W=w)}$ because $Pr(W=w,A=0) = Pr(W=w)Pr(A=0|W=w)$) in the second equality. 

To simplify further, note that the EIF is a unit-level function, so that summing a function involving $I(W=w)$ over $w$ simply pulls out that function evaluated at $W$ (the observed value for a given unit). Thus we have: 
    \begin{align*}
    EIF(\psi)&=  \frac{I(A=0)}{\Pr(A=0|W)}\bigg\{Y-q(W)\bigg\}  + q(W)-\sum_w q(w) p(w) \\
        &=\frac{I(A=0)}{g(W)}\bigg\{Y-q(W)\bigg\}  + q(W)-\psi
    \end{align*}
 In the second equality we substitute the notation $g(W)=\Pr(A=0|W)$, along with the definition of $\psi$.

It is important to reiterate that the above calculation only yields a function conjectured to be the true EIF, not necessarily the true EIF. For our purposes, the conjectured EIF is sufficient, since we go on to prove that our resulting estimator has the desired properties. If one needed to be sure this were the true EIF (e.g., to demonstrate efficiency for the proposed estimator), an additional proof would be required (see \cite{hines2022demystifying, kennedy2022semiparametric}). Typically, the true EIF will correspond to the conjectured EIF as derived above, though this is not always the case; for example, this approach will not work for a dose response curve of continuous treatment, as the latter does not have an EIF. The above approach also may not be sufficient if the estimand involves quantities other than conditional means and conditional probabilities; for example, quantile effects. Importantly, the EIF only exists if the estimand has a property called \textit{pathwise differentiability} (defined in Appendix \ref{app:eif}) meaning that the derivative of the mapping exists and has finite variance. A common violation of pathwise differentiability occurs when the above derivation leads to an expression containing an indicator for a variable that is in fact continuous (such as $I(W=w)$), and which cannot be further simplified. For example, the conditional ATE ($\E[Y^1 - Y^0|W=w]$) is not pathwise differentiable when $W$ is continuous since the resulting EIF derivation contains $I(W=w)$.
\section{Step 3: Derive an EIF-based estimator}
In Step 1 we equated the causal parameter to the statistical estimand $\psi$ under  identification assumptions. An intuitive (but not necessarily the best) choice of estimator for $\psi$ is the so-called \textit{plug-in estimator}, denoted $\psi ( \widehat P)$, where $\widehat P$ refers to estimators of the relevant pieces of the observed data distribution. In the case of our estimand, the plug-in estimator,  $\psi ( \widehat P)=\frac{1}{n}\sum_i \widehat q(W_i)$, corresponds to the g-computation estimator, and $\widehat P = \{ \widehat q(W), \widehat \Pr(W), \widehat g(W)\}$ contains estimators for the conditional outcome expectation, the empirical distribution of $W$, and the propensity score (though the plug-in estimator does not use the propensity score, we include it here for the benefit of notation below). Importantly, the plug-in estimator  $\psi ( \widehat P)$ does not generally allow for correct inference when the nuisance function $\widehat q(W)$ is estimated with machine learning; this drawback is addressed by using EIF-based estimators.

Here we illustrate how to create an EIF-based estimator of $\psi$. There are many types of estimators based on the EIF; here we present what is arguably the simplest, the so-called ``one-step'' estimator, which we denote $\widehat\psi_{os}$. Let $\phi(O_i, \widehat P)$ denote the estimated EIF; i.e. the EIF evaluated based on estimators $\widehat P$ of the relevant pieces of the observed data distribution, for unit $i$. Then, a one-step estimator is constructed by adding the sample mean of $\phi(O_i, \widehat P)$ to the plug-in estimator; i.e., $\widehat \psi_{os}=\psi(\widehat P) + \frac{1}{n} \sum_i \phi(O_i, \widehat P).$ (At this point, this form may seem arbitrary, but we provide some intuition in Section \ref{sec:proving_rate_dr} and more formal reasoning in Appendix \ref{app:von}).

For our example estimand, we obtain the following one-step estimator:
\small{
\begin{align*}
\widehat \psi_{os}&=\psi(\widehat P) + \frac{1}{n} \sum_{i=1}^n \left\{  \frac{\I(A_i=0) }{\widehat g(W_i)}\left(Y_i-\widehat q(W_i)\right) +  \widehat q(W_i)-  \psi(\widehat P)\right\} \\
   &= \frac{1}{n} \sum_{i=1}^n \left\{  \frac{\I(A_i=0) }{\widehat g(W_i)}\left(Y_i-\widehat q(W_i)\right) +  \widehat q(W_i)\right\}.
\end{align*}}%
\normalsize The above estimator has also been called the augmented inverse probability weighted (AIPW) estimator \cite{robins1997non}. To implement the AIPW estimator, we plug in estimates of $\hat g(W_i)$ and $\hat q(W_i)$ obtained from regressions (possibly fit using machine learning) of the treatment and outcome (respectively) on the covariates. Then we take the sample mean. 

\section{Step 4: Verify asymptotic properties of the estimator}
Just because an estimator is based on the EIF does not mean it can necessarily produce valid inference when machine learning is used; this requires proof of asymptotic properties. In particular, a property called \textit{rate double robustness} allows many EIF-based estimators to accommodate machine learning. Heuristically, rate double robustness means that the estimation error is a product of the errors of two nuisance function estimators. This means that as the sample size gets large, the errors of a rate doubly robust estimator go to zero faster than for non-rate doubly robust estimators (such as g-computation and inverse probability weighted (IPW) estimators), for which the estimation error is the average error of the respective nuisance estimator. This faster convergence rate means that standard inferential theory (such as the central limit theorem) applies, allowing one to construct valid standard errors. Next, we more carefully define rate double robustness and illustrate a method to prove it for our estimator. 

\subsection{Rate double robustness and why it is needed}
To define rate double robustness, we first need to define the \textit{convergence rate} of an estimator. Suppose an estimator is asymptotically consistent; i.e., as the sample size becomes large, both the bias and variance get smaller and eventually disappear. This property does not indicate \textit{how fast} the bias and variance disappear, called the convergence rate. We say an estimator $\widehat \psi$ has a convergence rate of $\sqrt{n}$ (i.e., is $\sqrt{n}$-consistent) and is asymptotically normal if: 
\begin{align}\label{eq:rootn_consstency}
    \sqrt{n}(\widehat \psi - \psi) \xrightarrow{d} N(0, \sigma^2),
\end{align}
where $\xrightarrow{d}$ denotes convergence in distribution. Intuitively, if the error $\widehat \psi - \psi$ (composed of bias and variance) converges to a mean-zero distribution with finite variance when scaled by $\sqrt{n}$, then this error is disappearing proportionally to the rate at which $\sqrt{n}$ is growing; i.e., the convergence rate is $\sqrt{n}$. Alternatively, if the estimator converges at slower than $\sqrt{n}$ rate, the error will blow up when scaled by $\sqrt{n}$; in particular, the scaled distribution may have variance that is increasing with $n$ rather than stable at $\sigma^2$. Convergence at $\sqrt{n}$ rate is important because (a) most statistical theory underpinning methods for calculating confidence intervals (including the bootstrap and other resampling methods) rely on scaling the errors by $\sqrt{n}$, and (b) $\sqrt n$ is usually the fastest possible rate of convergence. 

The key challenge with using machine learning to estimate nuisance functions is that many popular machine learning algorithms (such as random forests and boosted trees) converge at rates slower than $\sqrt{n}$; for example, these algorithms may converge at rate $n^{1/4}$. When these slower-converging algorithms are plugged into estimators that are not based on the EIF (e.g., g-computation, IPW), the resulting estimator will usually inherit these slower rates; as a result, the asymptotic distribution of the estimator is usually unknown and no inferential procedure is guaranteed to be valid. However, EIF-based estimators can oftentimes be shown to be $\sqrt{n}$-consistent and asymptotically normal even when the nuisance functions converge at slower rates. Specifically, for estimators with two nuisance functions (such as the one-step estimator $\widehat \psi_{os}$) we say that an estimator is \textit{rate doubly robust} if the estimator's convergence rate is equal to the product of the convergence rates of two nuisance functions. For example, if both are $n^{1/4}$-consistent, the estimator's convergence rate is $\left(n^{1/4}\right)^2=\sqrt{n}$. Further, slower rates in one nuisance estimator can be traded for faster rates in the other, as long as the product is $\sqrt{n}$. 

Note that rate double robustness is distinct from doubly robust consistency; the latter property (often simply called double robustness \cite{daniel2014double}) states that the estimator is consistent if one (but not necessarily both) nuisance estimators are correctly specified. In Appendix \ref{app:dr_consistency} we provide a formal definition of doubly robust consistency and a proof of this property for $\widehat \psi_{os}$.

\subsection{Proving rate double robustness}\label{sec:proving_rate_dr}  
It is important to restate that general methods for proving rate double robustness are complex, and our goal here is to provide introductory material summarizing the key concepts. For a deeper introduction, see \cite{hines2022demystifying,kennedy2022semiparametric}. 

To prove that an estimator is rate doubly robust, our goal is to show that it is $\sqrt n$-consistent even if the nuisance estimators are not. To show $\sqrt n$-consistency, we must determine whether the scaled error $ \sqrt n( \widehat \psi_{os}  - \psi (P) )$ converges asymptotically to a mean-zero distribution with bounded variance, as in equation (\ref{eq:rootn_consstency}). We begin this process by focusing on problems with the plug-in estimator, and then illustrate how the one-step estimator can address these problems. Specifically, the following is an expression for the error of the plug-in estimator, $\widehat \psi(\widehat P)  - \psi (P)$, known as the \textit{von Mises expansion}:
\begin{align}
\label{eq:firstorder} \psi (\widehat P) - \psi (P) &= -\underbrace{E\{\phi(O, \widehat P)\}}_{\text{``First order error''}} - \underbrace{R}_{\text{``Remainder''}}
\end{align}
In other words, the  error of the plug-in estimator can be decomposed into a ``first order error'' term which contains the EIF, and a ``remainder'' term. The derivation of this expression is outside our scope (see \cite{hines2022demystifying} for more detail). For intuition, it is helpful to know that (\ref{eq:firstorder}) is a first-order Taylor expansion; an expression involving the EIF shows up because the EIF is a first-order derivative of the parameter mapping.

For the plug-in estimator to be  $\sqrt n$-consistent, it would need to be the case that, for both terms on the right hand side of (\ref{eq:firstorder}), after scaling by $\sqrt n$, each term either disappears asymptotically or converges to a stable mean-zero distribution such as $N(0, \sigma^2)$. However, this will not generally be the case if a slower-converging machine learning estimator is used in $\widehat P$. This is because the ``first order error'' term inherits the slower rates and will blow up when scaled by $\sqrt n$. 

However, using an EIF-based estimator can directly address the first order error term under fairly general conditions, which we briefly explain here  (a fuller explanation is provided in Appendix \ref{app:von}). Specifically, because the one-step estimator contains an expression for the mean of the estimated EIF, it removes part of the first order error by design. The remaining part will generally converge to a mean zero normal distribution under a very important condition: that the nuisance estimators are fit in a separate sample from the one used to evaluate the final estimator. This can be accomplished by using various forms of sample splitting, such as cross-fitting \cite{newey2018cross}; for this reason, sample splitting is generally recommended whenever machine learning-based estimators are used \cite{zivich2021machine}. 

On the other hand, using an EIF-based estimator does not automatically address the remainder term $R$, as this part generally requires somewhat more involved derivations to show convergence. This is where the rate double robustness property comes into play: we must prove (or assume) that the remainder term has $\sqrt n$-convergence rate even when the nuisance estimators converge more slowly. In Appendix \ref{app:fullderiv}, we show a full proof of this property for our example estimand; here, we provide a heuristic overview to build intuition. We begin by solving (\ref{eq:firstorder}) for the remainder term: 
\begin{align}\label{eq:remainder0}
\nonumber    R &=  \psi(P) - \psi(\widehat P) - \E\{ \phi(O, \widehat P)\} \\
     &= \psi(P) - \psi(\widehat P) - \E\left\{ \frac{I(A=0)}{\widehat g}(Y - \widehat q) + \widehat q - \psi(\widehat P)\right\},
\end{align}
where we use the notation $g\equiv g(W)$, $\widehat g\equiv \widehat g(W)$, $q\equiv q(W)$, and $\widehat q \equiv \widehat q (W)$.  Equation (\ref{eq:remainder0}) shows that the remainder term captures the remaining error after removing the first order error. After a few algebraic manipulations (shown in Appendix \ref{app:fullderiv}) we achieve:
\begin{equation}\label{eq:remainder}
    R=-\E\left\{\frac{1}{\widehat g}(g-\widehat g)(q - \widehat q)\right\}
\end{equation}
The form of this term is instructive: the remainder equals a product of errors of the two nuisance functions, multiplied by $1/\widehat g$. This means that, under a mild condition, if $g-\widehat g$ converges to zero at rate $a$, and $q-\widehat q$ converges to zero at rate $b$, then $R$ converges to zero at rate $a \times b$. The argument follows from the Cauchy-Schwarz inequality, which here states that $R\leq \sqrt{\E\{ (g-\widehat g)^2\}} \times \sqrt{\E\{ (q-\widehat q)^2\}}$. Therefore, our goal in analyzing the remainder term is to transform it into the form $c\widehat e_1\widehat e_2$ (or a sum of such terms), where $c$ is a constant that is bounded in probability, and $\widehat e_1$ and $\widehat e_2$ are errors of nuisance function estimators. Then, Cauchy-Schwarz can be applied to show the rate double robustness property. (The ``mild condition'', a type of positivity assumption, is that both $g$ and $\widehat g$ are bounded by $(\epsilon, 1-\epsilon)$ where $\epsilon$ is some constant greater than 0.)

\section{Step 5. Derive a variance estimator for the EIF-based estimators}\label{sec:variance}
So far, we discussed how to propose an estimator and show that valid standard errors can be constructed when machine learning is used, but we have not yet stated \textit{how} these will be constructed. 

In the previous section, we showed that, for a given EIF-based estimator $\widehat \psi$ constructed with sample splitting, if we can show that the remainder term equals a product of errors of two nuisance functions such that Cauchy-Schwarz can be applied, then it is generally the case that:
\begin{equation}\label{eq:convergence}
    \sqrt n (\widehat \psi - \psi ) \xrightarrow{d} N(0, \E[\phi(O, P)^2]) 
\end{equation}
In other words, as the sample size gets large, the scaled errors approach a mean zero normal distribution with variance equal to the mean of the squared EIF, divided by $n$. This latter quantity is the asymptotic variance of the one-step estimator. Thus, a consistent estimator of the asymptotic variance is $\widehat V(\widehat \psi_{os})=\frac{1}{n^2}\sum_i^n  \phi(O_i, \widehat P)^2$, the sample variance of the estimated EIF, divided by $n$. This variance estimator relies on both nuisance functions being consistently estimated, which may be reasonable when machine learning is used.
\section{Discussion}
It can be fairly accessible to derive an estimator based on a conjectured EIF and to show that this estimator can accomodate machine learning. The calculations in this paper largely require only algebra, the sum and product rules for derivatives, and elementary knowledge of statistical inference. However, much care (and collaboration with a statistician) is required to provide rigorous proofs. It is also important to note that the rate double robustness property allows many, but not all, machine learning algorithms. The requirement is specifically that the convergence rates of the two algorithms, when multiplied together, achieve $\sqrt n$ rate - for example, both may be $n^{1/4}$, or one may be $n^{1/8}$ if the other is at least  $n^{3/8}$, etc. The convergence rate of a machine learning algorithm will generally depend on the tuning parameters and features of the data, and a rate such as $n^{1/4}$ is often reasonable (e.g., random forests and neural networks can acheive these rates \cite{farrell2021deep}) but not guaranteed. An ongoing area of work is to characterize rates of convergence for various algorithms \cite{chernozhukov2018double, farrell2021deep, chi2022asymptotic}. 

We focused on deriving rate doubly robust estimators, and emphasized that one should generally use sample splitting with these estimators so that one can use the most general types of machine learning algorithms. However, there are machine learning algorithms that can achieve fast enough rates that g-computation or IPW estimators can accommodate them, and some algorithms possess a property (known as Donsker class) that allows their use in rate doubly robust estimators without sample splitting. One fairly flexible machine learning algorithm that can achieve both of these properties (under the appropriate conditions) is the highly adaptive lasso (HAL) \cite{van2019causal}. 

Though we focused on one-step estimators, there are several other types of EIF-based estimators which can be shown to accommodate machine learning through similar means. These include estimating equation-based approaches, targeted maximum likelihood / minimum loss estimation, and double-debiased machine learning, which includes the one-step estimator as a special case \cite{hines2022demystifying, chernozhukov2018double}. The form of the variance estimator presented in Section \ref{sec:variance} can also be used for other EIF-based estimators. 

We hope this paper will empower epidemiologists and applied researchers to develop and apply estimators for identified causal estimands that are motivated by, and in service to, the research questions they actually want to answer, while avoiding unnecessary assumptions.
\newpage
\printbibliography

\newpage
\begin{appendix}
\section{Defining the EIF and pathwise differentiability}\label{app:eif}
To formally define the EIF, we must introduce \textit{parametric submodel}. Suppose the data are distributed according to the true distribution $P$, and consider another distribution $\widehat P$ whose support is contained in the support of $P$. A parametric submodel $P_e$ is a model that indexes a perturbance of $P$ in the direction of $\widehat P$ according to a finite dimensional mixture parameter, $e$. Typically $e$ is one-dimensional and a common choice is:
\[
P_e = e\widehat P + (1-e)P
\]
with $e \in [0, 1]$. Next we define the pathwise derivative for a mapping $\psi(P)$ as
\[
\lim_{e\downarrow 0}\left\{ \frac{\psi(P_e)-\psi(P)}{e}\right\} = \left.\frac{d\psi(P_e)}{de}\right|_{e=0}
\]
We say that the estimand $\psi$ is \textit{pathwise differentiable} when, for all regular parametric submodels (see \cite{hines2022demystifying} for a definition) the above derivative exists and can be written as:
\[
\left.\frac{d\psi(P_e)}{de}\right|_{e=0} = \int_{\mathcal O} \phi(o, P)\{d\widehat P(o) - d P(o)\},
\]
where $\phi(O, P)$ is mean zero and has finite variance (we use $\mathcal O$ to denote the support of $O$). We then formally define $\phi(O, P)$ as the efficient influence function for $\psi$. Note that because $\phi(O, P)$ has mean zero when evaluated at the true distribution $P$, the above quantity reduces to:
\[
\left.\frac{d\psi(P_e)}{de}\right|_{e=0} = \int_{\mathcal O} \phi(o, P)d\widehat P(o)
\]
In words, the derivative of the mapping applied to the parametric submodel, with respect to $e$, is the average of the efficient influence function, taken over the (wrong) distribution $\widehat P$.  This is technically what is meant by our statement that the efficient influence function is a type of derivative, and explains why derivative rules can be applied to conjecture influence functions.
\section{Formally analyzing the first-order error using the full von Mises expansion}\label{app:von}
Following the main text, we use $P_n$ to denote the sample mean operator, so that $P_nf=\frac{1}{n}\sum_{i=1}^nf(O_i)$ indicates the sample mean of $f(O)$, and $\widehat P$ to denote estimates of the relevant pieces of the distribution.
Following (\ref{eq:firstorder}), one may apply some additional algebra to achieve:
\begin{align}
    \label{eq:vonmises}
  \nonumber \sqrt{n}\left\{\psi(\widehat P) - \psi(P) \right\}&= \underbrace{-\sqrt n \E\{\phi(O, \widehat P)\}}_{\text{``First order error''}} -\sqrt{n}R \\
  &=\underbrace{\sqrt{n}P_n \{ \phi(O, P)\}}_{\text{``CLT term''}} - \underbrace{\sqrt{n}P_n \{\phi(O, \widehat P)\}}_{\text{``Drift term''}} + \underbrace{\sqrt{n}(P_n - \E)\{ \phi(O, \widehat P)-  \phi(O, P)\}}_{\text{``Empirical process term''}}  - \sqrt{n}R
\end{align}
The above is very commonly used decomposition, and in practice the von Mises expansion may refer to either the first or second equality. In (\ref{eq:vonmises}) we use the notation $(P_n - \E)f=P_n(f)-\E(f)$ to denote the difference of the empirical and true mean applied to the function $f(O)$ of the observed data. The expanded form of (\ref{eq:vonmises}) allows a more complete analysis. Since EIFs have mean zero by definition, the CLT term converges to a mean-zero normal distribution with finite variance. Replacing $\psi(\widehat P)$ with $\widehat \psi_{os}$ on the LHS above removes the drift term. The empirical process term converges in probability to zero when (i) $\widehat P$ is estimated in an independent sample from the one in which $\widehat \psi_{os}$ is calculated (which can be accomplished using cross-fitting \cite{chernozhukov2018double} or similar schemes), (ii) $\phi(O, \widehat P)$ converges to $\phi(O, P)$, which is often reasonable if machine learning is used; and (iii) $P_n$ converges to $P$, a mild condition that is usually true in practice. Thus, if we show that $\sqrt n R$ converges to $0$ under a certain set of conditions, this suffices to show that the above scaled error converges in distribution to a mean zero normal distribution with finite variance, allowing for inference based on the CLT. 
\section{Full derivation of the remainder term for the example estimand}\label{app:fullderiv}
First we introduce some additional notation. We use $o_p(a)$ to denote a random variable that is asymptotically equal to $0$ with convergence rate $a$. As above, we use $P_n$ to denote the sample mean, and $\widehat P$ to denote estimates of the relevant pieces of the distribution. we adopt the following:
\begin{assumption}[Positivity]\label{asn:pos1}
    For some constant $\epsilon > 0$ , $\Pr[\epsilon < \widehat g(W) < 1-\epsilon ]=\Pr[\epsilon < g(W)< 1-\epsilon ]=1$.
\end{assumption}
\begin{assumption}[Consistent estimation of nuisance functions]\label{asn:cons} $\widehat g \xrightarrow{p}g$ and $\widehat q \xrightarrow{p}q$, where $\xrightarrow{p}$ denotes convergence in probability.
\end{assumption}
Rerranging (\ref{eq:remainder0}), we have:
\begin{align}
\nonumber    R &= \psi(P) - \psi(\widehat P) - \E\left\{ \frac{I(A=0)}{\widehat g}(Y - \widehat q) + \widehat q - \psi(\widehat P)\right\} \\
\nonumber    &= \psi(P) -  \E\left\{ \frac{I(A=0)}{\widehat g}(Y - \widehat q) + \widehat q \right\}\\
\nonumber    &= \E\left\{ \frac{I(A=0)}{g}(Y - q) + q \right\} -  \E\left\{ \frac{I(A=0)}{\widehat g}(Y - \widehat q) + \widehat q \right\} \\
     &= \E\left\{ \frac{I(A=0)}{g}(Y - q) -\frac{I(A=0)}{\widehat g}(Y - \widehat q)\right\} + \E\left\{q - \widehat q \right\} \label{vm}
\end{align}
In the third equality we use $\psi(P)=\E[q]$ and the fact that $\E\left[\frac{I(A=0}{g}(Y-q) \right]=\E\left[\frac{I(A=0}{g}(q-q) \right]=0$ by law of iterated expectation.

First we analyse the $\E\left\{ \frac{I(A=0)}{g}(Y - q) -\frac{I(A=0)}{\widehat g}(Y - \widehat q)\right\}$ term:
\begin{align*}
  \E \left\{ \frac{I(A=0)}{g}(Y - q) -\frac{I(A=0)}{\widehat g}(Y - \widehat q)\right\} &=\E\left\{ \frac{I(A=0)}{g}(q - q) -\frac{I(A=0)}{\widehat g}(q - \widehat q)\right\} \\
  &=-\E\left\{\frac{g}{\widehat g}(q - \widehat q)\right\} \\
   &=-\E\left\{\frac{g}{\widehat g}(q - \widehat q) + \frac{g}{g}q - \frac{\widehat g}{\widehat g}q  + \frac{\widehat g}{\widehat g}\widehat q -  \frac{\widehat g}{\widehat g}\widehat q\right\} \\
   &=-\E\left\{\frac{g}{\widehat g}(q - \widehat q) - \frac{\widehat g}{\widehat g}(q - \widehat q) + q - \widehat q\right\} \\
    &=-\E\left\{\frac{g-\widehat g}{\widehat g}(q - \widehat q) + q-\widehat q \right\} ,
\end{align*}
where the first equality follows by law of iterated expectation, the second dropping terms equal to 0, the third by adding and subtracting 1 (twice), and the fourth and fifth rearranging. Then, plugging the above into (\ref{vm}):
\begin{align}\label{eq:remainder_ate}
   R = -\E\left\{\frac{g-\widehat g}{\widehat g}(q - \widehat q) + q-\widehat q\right\} + \E\left\{q-\widehat q \right\} =  -\E\left\{\frac{g-\widehat g}{\widehat g}(q - \widehat q)\right\} 
\end{align}
Under Assumption \ref{asn:pos1}, $1/\widehat g$ is bounded in probability. Thus, by the Cauchy-Schwarz inequality, it follows that $R \leq \sqrt{||g-\widehat g||}\sqrt{ ||q-\widehat q||}, $ where $||x||=\int x^2 dF(x)$ is the $L_2$ norm operator. This implies that $R = o_p(n^{1/2})$, so long as $g-\widehat g=o_p(n^a)$ and $q-\widehat q = o_p(n^b)$ where $a + b = 1/2$.

\section{Doubly robust consistency for the example estimand}\label{app:dr_consistency}
\subsection{Differentiating doubly robust consistency from rate double robustness}
Formally, we say that the estimator $\widehat\psi_{os}$ is doubly robust consistent if it converges in probability to (i.e., is consistent for) $\psi$ when either $\widehat q$ is consistent for $q$ or $\widehat g$ is consistent for $g$. In contrast, we say that the estimator $\widehat\psi_{os}$ is rate doubly robust if (i) both $\widehat q$ and $\widehat g$, and (ii) the convergence rate of $\widehat\psi_{os}$ equals the product of converge rates of $\widehat q$ and $\widehat g$. We provide a formal proof of doubly robust consistency for $\widehat\psi_{os}$ in the next subsection. 

Though the two properties are distinct, they are closely related, and doubly robust consistency can often be shown as a byproduct of the calculations used to show rate double robustness. In particular, from the analysis of the remainder term for $\widehat \psi_{os}$ in equation (\ref{eq:remainder}), it is apparent that terms involving $g-\widehat g$ will disappear asymptotically if $\widehat g$ is consistent for $g$, and likewise for $q-\widehat q$ if $\widehat q$ is consistent. If either of these terms goes to zero, the entire remainder term does.

Note that the variance estimator shown in Section \ref{sec:variance} relies on both nuisance functions being correctly specified. It is also possible to derive variance estimators which are themselves doubly robust consistent; i.e., they produce correct confidence interval coverage when only one nuisance function is correctly specified \cite{shook2024double}, though these are somewhat more complex.

\subsection{Proving doubly robust consistency}
Let $g^*$ denote the probability limit of $\widehat g$, and $q^*$ denote the probability limit of $\widehat q$, as $n\rightarrow \infty$. Our goal is to show that, if either $g^*=g$ or $q^*=q$ (not necessarily both), then for any $\epsilon > 0$,
\[
\lim_{n\rightarrow \infty} \Pr(| \widehat \psi_{os} - \psi | > \epsilon)=0 
\]
i.e., the one step estimator converges in probability to the statistical estimand. we use the von Mises expansion with both sides divided by $\sqrt{n}$, i.e.:
\begin{align}
    \label{eq:vonmises_norootn}
  \nonumber \psi(\widehat P) - \psi(P)&= \underbrace{P_n \{\phi(O, P)\}}_{\text{``CLT term''}} - \underbrace{P_n\{\phi(O, \widehat P)\}}_{\text{``Drift term''}}+ \underbrace{(P_n - \E)\{ \phi(O, \widehat P_n)-  \phi(O, P)\}}_{\text{``Empirical process term''}}  - R
\end{align}
Replacing $\psi(\widehat P)$ with $\widehat \psi_{os}$ on the LHS removes the drift term. Assuming that $P_n$ converges to $P$, the CLT term converges in probability to $0$. Likewise, the empirical process term converges in probability to $0$ since for any function $h(O)$ of the observed data, $(P_n - P)h(O)$ converges in probability to $0$. We are left show what $R$ converges to. By Slutzky's theorem, referring to (\ref{eq:remainder_ate}), it is clear that $R$ converges in probability to $0$ if either (i) $g-\widehat g$ or (ii) $q-\widehat q$. But, (i) will be the case if $g^*=g$, and (ii) will be the case if $q^*=q$. Finally, from the sum rule of limits, the entire left side converges in probability to $0$ if each summed element does.
\section{An alternative estimand: The expected untreated outcome among the treated}\label{app:alt}
In this section, we consider the estimand $\theta=\E[ \E(Y|A=0, W)|A=1]$, which equals $\E[Y^0|A=1]$ under conditional exchangeability, positivity, and causal consistency, as defined in Section \ref{sec:estimand}. For the estimand $\theta$, we conjecture an EIF, propose a one-step estimator, and show that the latter has the rate doubly robust property.

Note that the latter causal parameter is one piece of the average treatment effect in the treated, $\E[Y^1-Y^0|A=1]$. 
\subsection{Conjectured EIF and one-step estimator}
we use the notation $p(w, 1)\equiv \Pr(W=w|A=1)$.
    \begin{align*}
        EIF(\theta)&=EIF\left\{\sum_w q(w)p(w,1) \right\}\\
        &=\sum_w EIF\left\{q(w)p(w,1) \right\}\\
        &=\sum_w EIF\left\{q(w)\right\} p(w,1) +  \sum_w q(w) EIF\left\{p(w,1)\right\}\\
        &=\sum_w \frac{\I(W=w,A=0)}{\Pr(W=w,A=0)}\left[Y-q(w)\right]p(w, 1)  +  \sum_w q(w) \frac{\I(A=1)}{\Pr(A=1)}\left[\I(W=w)-p(w, 1)\right] \\
        &=\sum_w \frac{\I(W=w,A=0)}{\Pr(A=1)}  \frac{1-g(w)}{g(w)}\left[Y-q(w)\right]  + \sum_w \frac{\I(A=1)}{\Pr(A=1)} \left[q(w)I(W=w)-q(w)p(w, 1)\right]  \\
        &= \frac{\I(A=0)}{\Pr(A=1)}  \frac{1-g(W)}{g(W)}\left[Y-q(W)\right]  +  \frac{\I(A=1)}{\Pr(A=1)} \left[q(W)-\theta \right]
    \end{align*}
In the first equality, we pretend the data are discrete. In the second, we use the sum rule of derivatives, and in the third we use the product rule of derivatives. In the fourth equality, we substitute known EIFs of conditional means, in the fifth equality we apply Bayes rule, and in the sixth we use the fact that summing a function of $w$ involving
I(W = w) over w pulls out that function evaluated at the observed value $W$, and substitute the definition of $\theta$. 

From the conjectured EIF of $\theta$, we can propose the following one-step estimator:
\begin{align*}
    \widehat \theta_{os}&=\theta(\widehat P) +
    P_n\{\phi'(O, \widehat P)\}\\
    &=\theta(\widehat P) + P_n\left\{ \frac{\I(A=0)}{P_n(A)}  \frac{1-\widehat g(W)}{\widehat g(W)}\left[Y-\widehat q(W)\right]  +  \frac{\I(A=1)}{P_n(A)} \left[\widehat q(W)-\theta(\widehat P) \right]\right\}\\
    &= P_n\left\{ \frac{\I(A=0)}{P_n(A)}  \frac{1-\widehat g(W)}{\widehat g(W)}\left[Y-\widehat q(W)\right]  +  \frac{\I(A=1)}{P_n(A)} \widehat q(W)\right\}
\end{align*}
where we use $\phi'$ to denote the EIF of $\theta$, so that $\phi'(O, \widehat P)$ is the empirical EIF. In the second equality we use the fact that $P_n\left\{ \frac{\I(A=1)}{P_n(A)} \theta(\widehat P) \right\}=\frac{P_n(A)}{P_n(A)} \theta(\widehat P)$ since $P_n(A)$ and $\theta(\widehat P)$ are constant given the data.
\subsection{Proving rate double robustness}
From the previous subsection, we have that the conjectured EIF of $\theta$ is:
\begin{align}
    \phi'(O, P)&= \frac{\I(A=0)}{\Pr(A=1)}  \frac{1-g}{g}\left[Y-q\right]  +  \frac{\I(A=1)}{\Pr(A=1)} \left[q-\theta(P) \right].
\end{align}
Then, from (\ref{eq:firstorder}) we have:
\begin{alignat}{3}
    R&=\theta(P)-\theta(\widehat P)+\E[\phi(\widehat P)] \notag\\
    &=\E\left[\frac{I(A=0)}{P_n (A)}\frac{1-\widehat g}{\widehat g}(Y-\widehat q) + \frac{I(A=1)}{P_n(A)}[\widehat q - \theta(\widehat P)]   \right] + \theta(P)-\theta(\widehat P) \notag\\
    &= \E\left[\frac{g}{\widehat g}\frac{1-\widehat g}{P_n (A)}(q-\widehat q) + \frac{1-g}{P_n(A)}\widehat q   \right] - \frac{\Pr(A=1)}{P_n(A)}\theta(\widehat P)+ \theta(P)-\theta(\widehat P) \notag\\
    &= \E\left[\left\{\frac{g}{\widehat g} - 1 \right\}\frac{1-\widehat g}{P_n (A)}(q-\widehat q) + \frac{1-\widehat g}{P_n (A)}(q-\widehat q)+ \frac{1-g}{P_n(A)}\widehat q   \right] - \frac{\Pr(A=1)}{P_n(A)}\theta(\widehat P)+ \theta(P)-\theta(\widehat P) \notag\\
    &= \E\left[\left\{\frac{g}{\widehat g} - 1 \right\}\frac{1-\widehat g}{P_n (A)}(q-\widehat q) + \frac{1-\widehat g}{P_n (A)}(q-\widehat q)+ \frac{1-g}{P_n(A)}(\widehat q - q) + \frac{1-g}{P_n(A)}q   \right] - \frac{\Pr(A=1)}{P_n(A)}\theta(\widehat P)+ \theta(P)-\theta(\widehat P) \notag\\
     &= \E\left[\left\{\frac{g}{\widehat g} - 1 \right\}\frac{1-\widehat g}{P_n (A)}(q-\widehat q) + \frac{1-\widehat g}{P_n (A)}(q-\widehat q)+ \frac{1-g}{P_n(A)}(\widehat q - q)   \right] + \frac{\Pr(A=1)}{P_n(A)}[\theta(P) -\theta(\widehat P)]+ \theta(P)-\theta(\widehat P) \notag\\
    &=\E\left[\frac{g-\widehat g}{\widehat g} \frac{1-\widehat g}{P_n (A)}(q-\widehat q)\right] \label{s1}\\
    &\quad + \E\left[ \frac{\widehat g - g}{P_n (A)}(\widehat q-q)\right] \label{s2}\\
    &\quad - \frac{\Pr(A=1) - P_n(A)}{P_n(A)}\left[\theta(P) -\theta(\widehat P)\right].\label{s3}
\end{alignat}
In the third equality, we use iterated expectation and the fact that $P_n(A)$ and $\theta(\widehat P)$ are constants. In the fourth equality, we add and subtract $\frac{1-\widehat g}{P_n (A)}(q-\widehat q)$. In the fifth equality, we add and subtract $\frac{1-\widehat g}{P_n (A)}q$. In the sixth equality, $\E\left[ \frac{1-g}{P_n(A)}q \right]=\E\left[ I(A=1){P_n(A)}q \right]=\E\left[ \frac{I(A=1)}{P_n(A)}\E[q|A=1] \right]=\frac{\Pr(A=1)}{P_n(A)}\theta(P)$ because $P_n(A)$ and $\E[q|A=1]$ are constants and the latter equals $\theta(P)$. In (\ref{s1}) we use $1=\frac{\widehat g}{\widehat g}$, in (\ref{s2}) we use $1-\widehat g - (1-g)=g-\widehat g$, and in (\ref{s3}) we use $1=\frac{P_n(A)}{P_n(A)}$. 

Then it is straightforward to see that each of (\ref{s1}-\ref{s3}) has the rate double robust property. By Assumption \ref{asn:pos1}, we have that the following are bounded in probability:  $ \frac{1}{\widehat g} \frac{1-g}{P_n (A)}$ (appearing in (\ref{s1})) and $\frac{1}{P_n(A)}$ (appearing in (\ref{s2}) and (\ref{s3})). Then it follows that (\ref{s1}) and (\ref{s2}) converge to zero at $\sqrt n$ rate if the product of convergence rates of $\widehat g$ and $\widehat q$ is $\sqrt n$. Since $P_n (A)$ is a sample mean, by the central limit theorem it converges to $\Pr(A=1)$ at $\sqrt n$ rate. Thus, as long as $\theta(\widehat P)$ is consistent for $\theta(P)$ (which is guaranteed by Assumption \ref{asn:cons}), then (\ref{s3}) will always converge to zero at $\sqrt n$ rate. This illustrates the dependence of the rate double robustness property on consistent estimation of the nuisance functions: consistency of the plug-in estimator $\theta(P)$ is needed to ensure (\ref{s3}) converges at the right rate. Finally, if (\ref{s1}), (\ref{s2}), and (\ref{s3}) are all converging at $\sqrt n$, their sum will as well.
\end{appendix}
    
\end{document}